\def\lae{\mathrel{<\kern-1.0em\lower0.9ex\hbox{$\sim$}}}
\def\gae{\mathrel{>\kern-1.0em\lower0.9ex\hbox{$\sim$}}}

\newcommand{\be}{\begin{equation}}
\newcommand{\ee}{\end{equation}}

\documentclass{emulateapj}

\slugcomment{Accepted for publication in ApJ Letters}
\lefthead{JORD\'AN }
\righthead{GC HALF-LIGHT RADII}

\begin{document}

\title{
A Possible Explanation for the Size Difference
of Red and Blue Globular Clusters}

\author{Andr\'es Jord\'an\altaffilmark{1,2,3}}

\begin{abstract}
Most observations of the projected half-light radii of 
metal-rich globular clusters in a variety of galaxies
have shown them to be $\sim 20\%$
smaller than those of their metal-poor counterparts.
We show using multi-mass isotropic Michie-King models
that the combined effects of mass segregation
and the dependence of main sequence lifetimes on metallicity
can account for this difference, under the assumption
that clusters with similar central potentials
have the same distribution of half {\it mass} radii. 
If confirmed, this would represent a new constraint
on theories of globular cluster formation and evolution.
\end{abstract}


\altaffiltext{1}{Department of Physics and Astronomy, Rutgers University,
136 Frelinghuysen Rd, Piscataway, NJ 08854, USA}
\altaffiltext{2}{Claudio Anguita Fellow}
\altaffiltext{3}{Astrophysics, Denys Wilkinson Building, University of Oxford,
1 Keble Road, Oxford, OX1 3RH, UK; andresj@astro.ox.ac.uk}

\section{Introduction}

The last decade has seen rapid progress in the
characterization of globular cluster (hereafter GC)
systems in external galaxies.
An important task is to disentangle the properties of GCs
which are universal form those that are correlated
with other of their properties or those of their host galaxies.
The {\it Hubble Space Telescope} can partially resolve
the spatial profiles of GCs well beyond the Local Group,
and thus study their structural parameters.
The recovery of these parameters
requires modeling of the point-spread function, 
and this has been carried out using different methods 
and instruments by various groups, using a range
of galaxies including spirals and ellipticals
(Kundu \& Whitmore 1998; Kundu et~al. 1999; Puzia et~al. 
1999; Larsen, Forbes \& Brodie 2001; Barmby, Holland \& 
Huchra 2002; Larsen et~al. 2001; Harris et~al. 2002; 
Jord\'an et~al. 2004). Most of these studies have revealed
that metal-rich (red) GCs appear to have half-light radii
$\sim 20\%$ smaller than their metal-poor (blue) counterparts.

Larsen \& Brodie (2003) have advanced the only plausible
explanation so far for this size difference. They argue
that the observed difference can arise as a projection effect,
resulting from combination
of a correlation between galactocentric distance
and size 
and the differing spatial
distributions of the GC subpopulations. Assuming that
GCs in all galaxies follow a relation between galactocentric 
distance and size similar to that of Galactic GCs, they find
that this mechanism is able produce the observed size 
difference, albeit
with some fine tuning.

In this Letter we propose a simple explanation for
the observed difference. We propose that the difference
is a consequence of mass segregation and 
the fact that lower metallicity stars 
have longer lifetimes for a given mass. 
Assuming that the average half {\it mass} radius
does not depend on metallicity, 
we model the observed light profiles
with Michie-King multi-mass models and stellar isochrones,
and show that a size difference of the observed
magnitude arises naturally.

\section{Models}

We model GCs using Michie-King multi-mass isotropic models.
The formalism is described in Gunn \& Griffin (1979). In what
follows, we repeat some of their expressions restricted to the
isotropic case to set the notation.

Each mass class is labeled by an integer $j$ and is assumed
to have a distribution function of the form $f(E) = e^{A_jE}-1$,
where $E=\frac{1}{2}v^2+\psi$ and $\psi$ is the potential energy. Due
to the short relaxation times at the core, we assume thermal
equilibrium there, which demands that $A_j = \beta \bar{m}_j$,
where $\beta$ is a constant and $\bar{m}_j$ is 
the mean mass of the $j$-th class. Choosing a characteristic
radius $r_c$ and velocity variance $v_0^2$, 
and letting $W=-\psi/v_0^2$ and $\xi=r/r_c$, the Poisson
equation reads $\nabla_{\xi}^2 W = -9\sigma$,
where $\sigma = \rho/\rho_0$, $\rho=\sum_j \rho_j$ is the
mass density and $\rho_0 \equiv \rho(0)$. 

If we define $\alpha_j=\rho_{0j}/\rho_0$ to be the fractional
density contribution of mass class $j$ at the center, a model
is completely specified by a value for $W$ at the center, $W_0$,
and the $\{\alpha_j\}$. With these quantities specified, 
the Poisson equation is integrated until $W=0$,
at which point the tidal limit is reached. 
The normalized mass densities for class $j$,
$\sigma_j(r)$, are then calculated.
The models have mass segregation, and there is
no expression relating the $\{\alpha_j\}$ to the {\it total}
mass in class $j$, which is the quantity we would like to specify
via a global initial mass function $\zeta(m)$, defined
such that the total number of stars in the GC with initial masses
between $m$ and $m+dm$ is $\zeta(m)dm$,
and a relation between initial
and present mass, $m_f(m)$. 
The problem is solved by 
iteration until the $\{\alpha_j\}$ and the total mass
in class $j$, $M_j = \int_j m_f(m) \zeta(m)\,dm / (\sum_i M_i)$,
are self-consistent. 
The space densities are finally projected to obtain the 
observed mass densities.

Determining the initial mass function in GCs is a difficult task
because of the need to take into account the effects of mass
segregation. Using Michie-King 
multi-mass models to account for dynamical effects, 
Paresce \& De Marchi (2000) 
find that the mass functions of a dozen globulars
are well described by a lognormal distribution with a mean
$m_c=0.33$ and dispersion $\sigma=0.34$. We will adopt this
as the mass function for GCs for $m \le 0.9M_{\odot}$. At higher masses,
it is found that the IMF is well described by a power law
with an exponent close to Salpeter (Chabrier 2003). Thus, we adopt
$\zeta (m) \propto m^{-x}$ for $m > 0.9M_{\odot}$. In what follows
we assume $x=2$, as this allows us to obtain a good match to the
observed size difference in M87. While this value is certainly
consistent with the observations, there is otherwise 
no fundamental reason for our choice. Below we comment
the effect of varying $x$.

We construct the mass classes as follows. We take the minimum
mass in the initial mass function 
to be $0.1 M_{\odot}$ and the maximum to be $30 M_{\odot}$. 
If $m_{to}$ is the mass at the main sequence turnoff, $m_{trgb}$ the
mass at the tip of the red giant branch and 
$N = \lceil (m_{to}-0.4)/0.1 \rceil \equiv (m_{to}-0.4)/\Delta m $, 
the limits of the mass classes are
$(0.1,0.2,0.3,0.4,0.4+\Delta m,\ldots,0.4+(N-1)\Delta m,
m_{to},m_{trgb},2,3,4,5,6,7,8,30)$. 
The masses in each class are obtained differently depending on
whether the value of the mass is greater than $m_{trgb}$. We ignore
in what follows the evolutionary stages between the tip
of the red giant branch and a white dwarf. This should not 
affect greatly the derived light profiles, as those stars will
have density profiles close to that of stars with $m_{to}$, and thus will
just add light to the already dominating component.
 For $m < m_{trgb}$, we have simply
$M_j \propto \int_j m \zeta(m) dm$. We assume that stars with 
$m_{trgb} < m < 8$ end their lives as white dwarfs and that stars 
with $8<m<30$ end as neutron stars with a mass of $1.4M_{\odot}$.
For the masses $m_{trgb} < m < 8$,
we determine the relation between the white dwarf mass and the star's 
initial mass, $m_{wd}(m)$, by interpolating linearly the data
presented in Table~3 of Weidemann (2000).
The mass of the white dwarf bins are given then by
$M_j \propto \int_j m_{wd}(m) \zeta(m) dm$. Neutron stars
are observed to be born with velocity kicks of hundreds of 
km $s^{-1}$ (e.g. Lyne \& Lorimer 1994), and so most of them should escape 
from their host GCs.
Of course, the presence of millisecond pulsars and low mass X-ray binaries in 
GCs means that some
of the neutron stars must be retained.
Here we will
assume that a fraction $f_{ns}=0.05$ of neutron stars are retained
by a typical GC.  
The mass in the neutron star mass class is then
$M_{ns}\propto 1.4 f_{ns} \int_8^{30} \zeta(m) dm$.
The adopted  value for $f_{ns}$ is consistent with the results
of a comprehensive study of neutron star retention in GCs by Pfhal, 
Rappaport \& Podsiadlowski (2002). 
Our results are
not very sensitive to the precise value adopted for $f_{ns}$
and remain 
essentially unchanged when varying $f_{ns}$ in the range $0.01-0.1$. 

The final ingredient to obtain the observed light profiles
is stellar isochrones, from which we obtain the mass-luminosity
function $L(m)$. We used the Y$^2$ isochrones 
(Yi et~al. 2003). Observational evidence
points to most GCs in early-type galaxies being an old and 
coeval population (see, {\it e.g.}, Jord\'an et~al. 2002).
We will thus assume that GCs have an age of 13 Gyr, the mean age 
inferred  for the GC system of M87 using spectroscopic line
indices (Cohen, Blakeslee \& Ryzhov 1998).
We interpolated stellar
populations with [Fe/H] in $\{-2,-1.75,-1.5,-1.25,-1,-0.75,-0.5,-0.25,0\}$,
an age of 13 Gyr and alpha enhancement $\alpha=0.3$. Note that 
the stellar
isochrones have a minimum mass of $0.4 M_{\odot}$. To obtain
the luminosities of lower mass stars we used the zero-age 
main sequence mass-luminosity relation of Tout et~al. (1996), 
ensuring continuity with the luminosity given in the Y$^2$ isochrones
for $m=0.4 M_{\odot}$. With a given isochrone in hand, we determined
$m_{to}$, $m_{trgb}$ and constructed the corresponding Michie-King
multi mass model. The projected densities of each of the mass classes
with $m < m_{trgb}$were then multiplied by the mean $V$-band luminosity
$L_{Vj}$ obtained from the isochrones as 
$L_{Vj} = \int_j L(m)\zeta(m) \,dm / \int_j \zeta(m)\,dm$.

\begin{figure}
\plotone{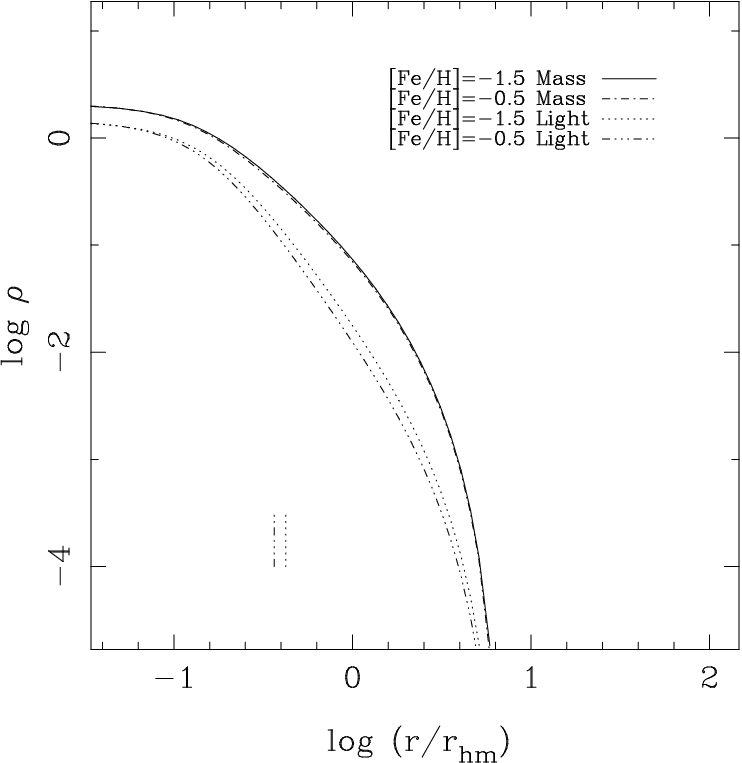}
\caption[]{Projected mass and light profiles for models
with $W_0=9$ and [Fe/H]$=-1.5$ and [Fe/H]$=-0.5$, typical
of metal-poor and metal-rich GCs respectively in 
early-type galaxies. The vertical lines indicate the half
light radii $r_{hl}$ in units of the half mass radius of the respective
model, showing
that the metal-rich model has a half light radius which
is $\sim 14\%$ smaller than that of its metal-poor counterpart.
\label{fig:models}
}
\end{figure}

As described above, the models constructed are specified 
by a value of $W_0$ and the set of $\{\alpha_j\}$. There remain
two arbitrary scale factors, which correspond to setting the
scale for the spatial coordinates and the overall mass of the 
system. In order to compare the models against each other, we set
the half {\it mass} radius $r_{hm}$ to the same physical length for 
all models. 
This assumption would follow
if the overall structure of young GCs is determined
by processes mostly independent of metallicity, and 
if GCs were subjected afterwards on average to the same dynamical 
effects. 
For each model, we then recorded the projected half {\it light}
radius $r_{hl}$ in units of $r_{hm}$ (for each metallicity and
$W_0$). 

\begin{figure}
\plotone{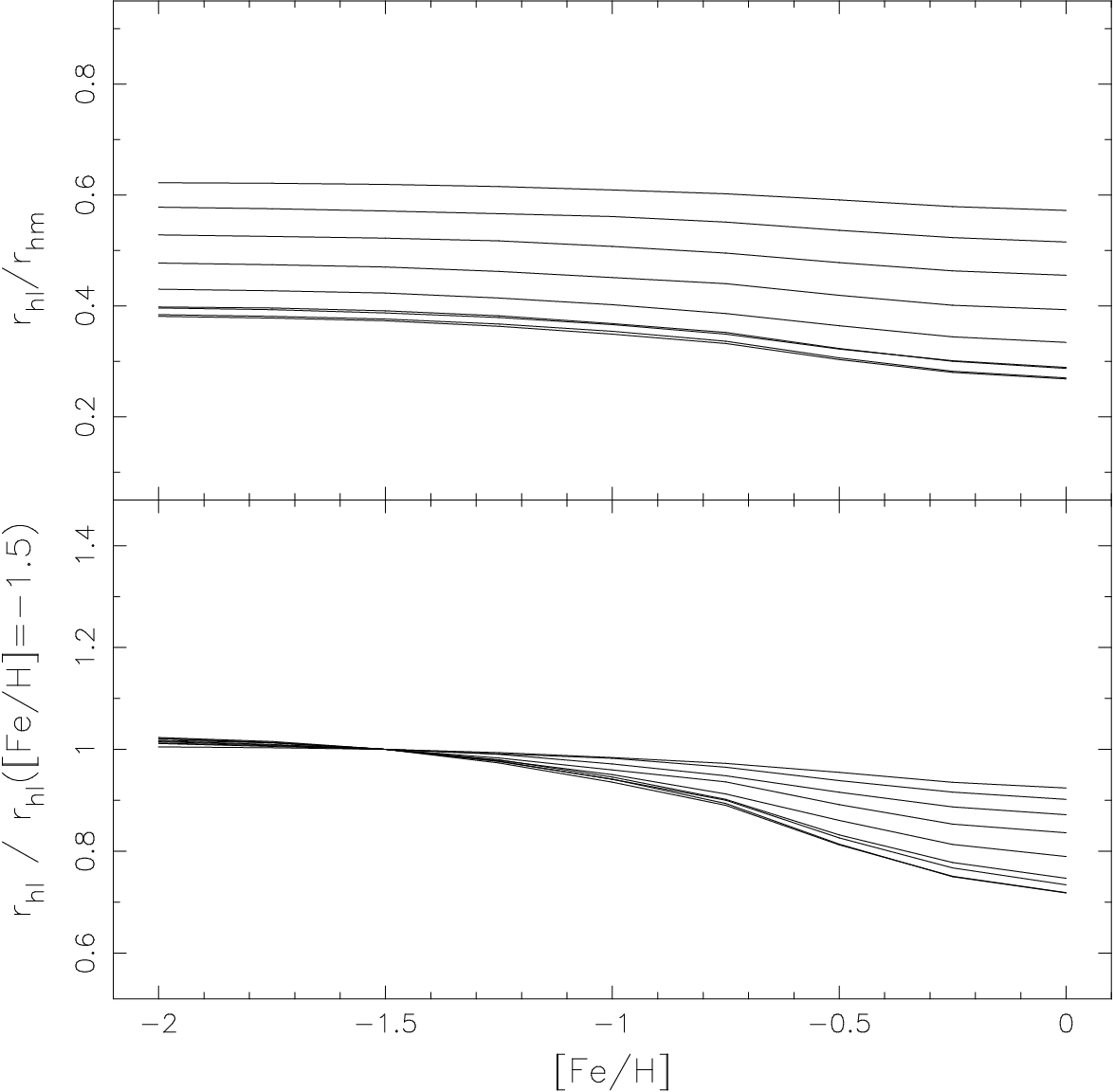}
\caption[]{ ({\it Top}) Projected half light radius $r_{hl}$ in units of 
the half
mass radius as a function of [Fe/H]. The different curves
correspond to different values of the central potential $W_0$,
ranging from $W_0=5$ (upper curve) to $W_0=13$ (lower curve)
in steps of $\Delta W = 1$.
({\it Bottom}) This panel shows the same set of curves normalized
to their value at [Fe/H]$=-1.5$, a typical metallicity for a metal-poor
GC. 
\label{fig:rh_feh}
}
\end{figure}

\section{Results and Discussion}

In Figure~\ref{fig:models} we show mass and light profiles
for two models with $W_0=9$ and [Fe/H] equal to $-1.5$ and $-0.5$. The half
light radii are indicated in the figure, and it can be seen that
$r_{hl}$ is smaller for the metal-rich model by $14\%$.
This is because the mass of the most luminous stars becomes
larger as the metallicity increases, and thus their density profile
is more concentrated.
In Figure~\ref{fig:rh_feh} we show $r_{hl}$ as a function
of [Fe/H] for various values of $W_0$. It is evident
from the figure that for a given value
of $W_0$, $r_{hl}$ gets smaller as the metallicity 
increases. Note that the half light radii
are roughly half the corresponding half-mass radii.
In order to see directly the size of the effect,
we also plot the same curves normalized to their values
at [Fe/H]$=-1.5$, typical of metal-poor GCs. 
A typical metal-rich GC in early-type galaxies will have 
[Fe/H] $\sim -0.3$. The size of the effect at that metallicity 
is in the range $\sim 5-25\%$, the exact value depending on the value
of $W_0$. We will take a $W_0=9$ model to be representative
of a typical GC, as the light profile of such a model will have
$\log (r_t/r_{cl})\sim 1.58$, where $r_t$ is the tidal radius and 
we have defined the
core radius of the projected light profiles $r_{cl}$ in a way 
akin to single mass King models, which for a concentration
$c\sim 1.5$ is $1.05$ times the radius at which the luminosity
is half the central value. 
Thus, this concentration measure is comparable to a typical 
concentration of a Galactic GC (Harris 1996).
From Figure~\ref{fig:rh_feh} we see that a typical metal-rich
GC in an early-type galaxy will be observed to be $\sim 20\%$ smaller than a
corresponding metal-poor GC. This is consistent with the 
observations, and thus our models can explain them
by assuming
that GCs have universal physical properties and the combined effects
of mass segregation and stellar evolution.

\begin{figure}
\plotone{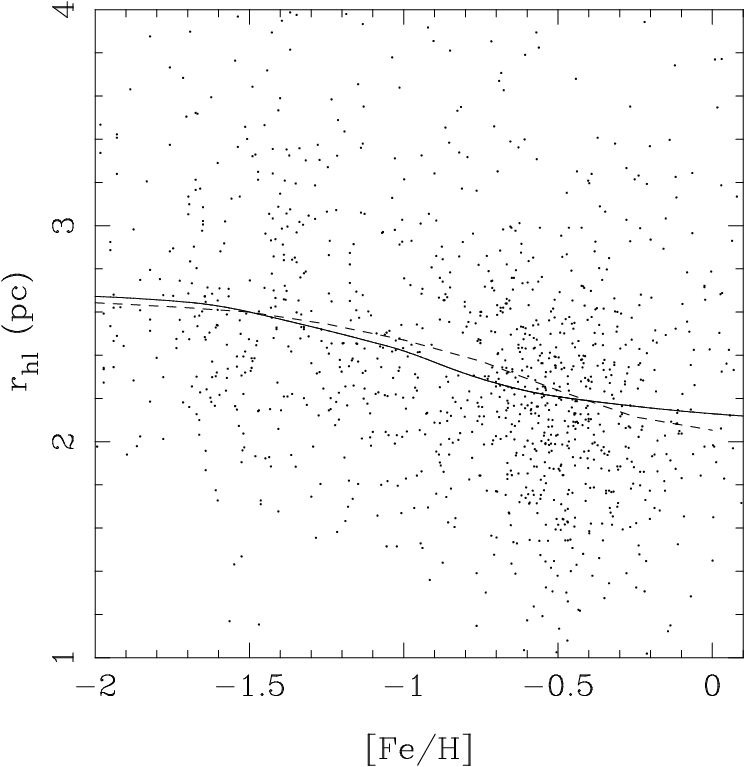}
\caption[]{ 
The dots are the projected half light radius for a sample of GCs in 
M87 measured using ACS images. The solid curve represents a robust estimate
of the mean half light radius as a function of [Fe/H], and the dashed
line is the predicted behavior of this quantity for a model
with $W_0=9$ which is normalized to the observed value at [Fe/H]$=-1.5$.
\label{fig:m87}
}
\end{figure}

We can go beyond the mean difference in size, as our procedure
gives a definite prediction for the behavior of the
half light radius as a function of [Fe/H]. The proper way of comparing
with the observations would be to input the distribution of $W_0$, 
and then get the predicted behavior for $r_{hl}$. 
As we don't have the distribution
of central potentials available to us, we will assume as above that
the population of GCs is well represented by a model with $W_0=9$.
In Figure~\ref{fig:m87} we show the measured half light radii
(average of $g_{475}$ and $z_{475}$ measurements) with uncertainties less
than 0.5 pc for GCs in M87,
using data from the ACS Virgo Cluster Survey 
(C\^ot\'e et~al. 2004). These measurements form part of a systematic
investigation of structural parameters for GCs in Virgo galaxies
(Jord\'an et~al., in preparation). 
The solid line is a robust 
smoothing of the data done with the Lowess (Cleveland 1979) method. 
While we show in the figure only GC candidates with $1$ pc $<r_{hl}<4$ pc,
there was no restriction in $r_{hl}$ for the analysis.
The values of [Fe/H] are obtained
from ($g_{475}-z_{475}$) as described in Jord\'an et~al. (2004).
The dashed line is the predicted behavior of our models for $W_0=9$,
where we have set the normalization such that the curves coincide at
[Fe/H]$=-1.5$. The agreement is very good, especially when considering
the crudeness of comparing with a single value of $W_0$.

The results above depend on the assumption that, given a value of $W_0$,
the average half mass radius does not depend on [Fe/H].
This assumption
is appealing in that it points to a universality in the
formation and evolution process of GCs. 
If clusters observed today with a certain central potential
were formed with the same average half mass radius and were subjected,
{\it on average}, to the same dynamical effects from 
the potential field of their galaxy and internal mass loss
processes, they should have on average the same
half mass radii. Individual GCs might of course been subjected
to quite different histories.
The process of GC formation is yet to be fully understood,
and thus there are few theoretical handles that would let us
assess the plausibility of assuming a constant average
half-mass radii with metallicity.
Some proposed formation mechanisms
determine the overall scale of the proto-GCs by mechanisms
that should be largely independent of the metal content,
such as cosmological reionization compression
of subgalactic halos (Cen 2001), formation out of dense
cores of supergiant molecular clouds (Harris \& Pudritz 1994)
or on the high mass and pressure clouds of gas
partitioned by supersonic turbulence (Elmegreen \& Efremov 1997).
Observationally, Larsen (2004) finds no evidence for
variations on the average sizes of young stellar clusters in a sample
of nearby spirals. Overall, the assumption seems certainly plausible
in light of our current understanding of GC formation. 

There are a number of factors that can affect the predicted behavior
of our models. An obvious one is the shape of the mass function,
as this will change the derived $\{\alpha_j\}$. As an example,
changing the power law exponent for the high mass part of 
our assumed mass function to $x=2.35$ reduces the magnitude of the difference 
by a factor of $\sim 0.86$. Another factor which should have an effect 
insofar as it will
change the evolutionary state of the stellar populations is age. 
If the populations are coeval, changing the age from 13 to 11 Gyr
reduces the magnitude of the difference by a factor of $\sim 0.86$.
If there is an age difference between the subpopulations 
they would have evolved dynamically for a different
total time and the assumption of them having the same half mass radius
would be less warranted.
Predicting the effect of varying the age properly would thus require knowledge
of how the average half mass radii evolves through dynamical effects.
This caveat notwithstanding, the fact that most models show $r_{hm}$
to be a rather stable quantity over the cluster's evolution 
(e.g., Aarseth \& Heggie 1998)
make it reasonable to use the present models and the assumption
of constant half-mass radius to get an estimate of the effect
of an age difference. If clusters with [Fe/H]$>-1$ are 
3 Gyr younger than their metal-poor counterparts, the magnitude
of the difference would be increased by a factor of $\sim 1.5$.
At any rate, most determinations
of the relative ages of GCs in early-type galaxies, where most of the size
difference observations have been made, 
are consistent with the GCs being roughly coeval
(see, e.g., Jord\'an et~al. 2002 and references therein). 
The relation
$m_{wd}(m)$ for white dwarf remnants also plays an important
role in determining the magnitude of the difference as they
contribute an appreciable fraction ($\sim 20\%$) to the cluster mass.

Variations of the distributions of $r_{hm}$ and $W_0$
with [Fe/H] or galactocentric radius can potentially be important, 
but the main point
to be made from our results is that the 
observations can comfortably be reproduced without
resorting to intrinsic differences in the sizes of the 
GCs as a function of metallicity. The use
of Michie-King multi mass models is well-suited
to this task, but for a more precise determination
of the expected size difference, and the dependence of 
it with variations in the input ingredients,
it would be very useful that this effect be 
followed with N-body simulations
that take into account the effects of stellar evolution and 
of the gravitational potential of the host galaxy 
(e.g., Baumgardt \& Makino 2003). 
With more detailed models in hand,
the dependencies of the predicted half light
radii could perhaps be used  with other 
observed properties to simultaneously constraint
factors such as age and the form of the initial mass
function. 

Our models let us predict that if the GCs
are roughly coeval, then the size of the observed average half light
radius difference should scale with the mean metallicity
of the metal-rich GCs. This is known to correlate with 
galaxy luminosity (Brodie \& Huchra 1991), 
so the average size difference should
scale with it. 
In contrast with the proposal of Larsen \& Brodie (2003)
our models do not predict a change in the size difference
with projected radius. Larsen \& Brodie (2003)
argue that some pointings away from the central
regions in M87 do not show a significant size difference
between the GC subpopulations.
Although a dilution of the size difference with radius 
could be introduced in our models by radial variations in 
other factor such as the mean [Fe/H] of the metal-rich 
subpopulation, it is unlikely that the effect will
disappear at large radii without some fine tuning. Thus,
a larger number of observations 
of size difference at large galactocentric radii will be very useful
in discriminating between the models. It is possible
that the overall effect results as a combination
of projection and mass segregation effects, the former
disappearing at large galactocentric radius.
We stress that large samples of GCs
are needed to investigate this issue, as GCs will have 
a distribution of intrinsic radii and we need an accurate
determination of the {\it average} behavior. So while
no size difference was reported in NGC~5128 by
Harris et~al. 2002, the low number of clusters 
they observed, as they note, precludes the drawing
of strong constraints. 

We suggest in light of the models
presented here that the size differences observed so far
are consistent with GCs having half mass radii distributions
that do not depend on metallicity.
If true, GCs would present us with another universal property,
such as the shape of their luminosity function (Harris 2001)
or their formation efficiencies (Blakeslee et~al. 1997; McLaughlin 1999),
which can hold an important clue to their
formation and subsequent evolution. 
And even if the effects
of mass segregation do not account entirely for the observed 
size difference,
it is clear that its contribution must be included
when interpreting the observations and their implications
for GC formation and evolution.

\acknowledgements
The author thanks Pat C\^ot\'e and Tad Pryor for useful 
discussions and comments
on the manuscript and the ACSVCS team for granting use of the M87
data in advance of publication. Support for program GO-9401 was provided
through a grant from the Space Telescope Science Institute, which is 
operated by the Association of Universities for Research in Astronomy, Inc.,
under NASA contract NAS5-26555.
Additional support was provided by the National Science 
Foundation 
through a grant from the Association of Universities for Research in 
Astronomy, Inc., under NSF
cooperative agreement AST-9613615 and by Fundaci\'on Andes under project 
No.C-13442.

\end{document}